\documentclass[aps,reprint,groupedaddress,superscriptaddress,prl]{revtex4-2}

\usepackage[utf8]{inputenc}
\usepackage{amsmath}
\usepackage{amsfonts}
\usepackage{amssymb}
\usepackage{bm}
\usepackage{color}
\usepackage{tcolorbox}
\usepackage{graphicx}
\usepackage{soul}
\usepackage{physics}
\usepackage{mathrsfs}
\usepackage{multirow}
\usepackage{xcolor}
\usepackage{mathtools}
\usepackage{lipsum}
\usepackage{comment}
\usepackage{booktabs}
\usepackage{textcomp}

\usepackage[colorlinks,citecolor=blue,linkcolor=red,anchorcolor=blue,urlcolor=blue]{hyperref}

\newcommand{\cA}[1]{\multicolumn{1}{c|}{\parbox{0.45\textwidth}{\centering #1}}}
\newcommand{\cB}[1]{\multicolumn{1}{c|}{\parbox{0.30\textwidth}{\centering #1}}}
\newcommand{\cC}[1]{\multicolumn{1}{c}{\parbox{0.20\textwidth}{\centering #1}}}

\begin{document}
\title{Coercivity Landscape Characterizes Dynamic Hysteresis}
        
\author{Miao Chen}
\altaffiliation{These authors contributed equally to this work.}
\affiliation{School of Physics and Astronomy, Beijing Normal University, Beijing, 100875, China}
\affiliation{Key Laboratory of Multiscale Spin Physics (Ministry of Education), Beijing Normal University,Beijing 100875, China}

\author{Xiu-Hua Zhao}
\altaffiliation{These authors contributed equally to this work.}
\affiliation{School of Physics and Astronomy, Beijing Normal University, Beijing, 100875, China}
\affiliation{Key Laboratory of Multiscale Spin Physics (Ministry of Education), Beijing Normal University,Beijing 100875, China}

\author{Yu-Han Ma}
\email{yhma@bnu.edu.cn}
\affiliation{School of Physics and Astronomy, Beijing Normal University, Beijing, 100875, China}
\affiliation{Key Laboratory of Multiscale Spin Physics (Ministry of Education), Beijing Normal University,Beijing 100875, China}
\affiliation{Graduate School of China Academy of Engineering Physics, Beijing, 100193, China}

\begin{abstract}Hysteresis, with rich dynamical behaviors—especially in interacting systems—has drawn broad research interest. Yet its dynamic scalings across time scales lack a unified description, and their transitions remain unclear. Here, we study the stochastic $\phi^4$ model driven periodically by an external field $H$. For large systems with small noise strength $\sigma$, we find the coercivity $H_c \equiv H(\langle\phi\rangle=0)$ sequentially exhibits distinct behaviors with increasing  driving rate $v_H$: $v_H$-scaling increase, stable plateau ($v_H^0$), $v_H^{1/2}$-scaling increase, and abrupt decline to disappearance. The plateau reflects the competition between thermodynamic and quasi-static limits, namely, $\lim_{\sigma\to 0}\lim_{v_H\to 0}H_c = 0$, and $\lim_{v_H\to 0}\lim_{\sigma\to 0}H_c=H^*$. Here, $H^*$ is exactly the field-driven first-order phase transition point. In the post-plateau regime, $(H_{c} - H_{P})$ scales with $(v_{H} - v_{P})^{2/3}$ with $v_{P}$ and $H_{P}$ being the reference points of the plateau. Moreover, we reveal a finite-size scaling for the coercivity plateau as $v_{P}\sim\sigma^{2}$ and $(H^*-H_P)\sim\sigma^{4/3}$ by utilizing renormalization-group theory. Our work provides a panoramic view of finite-time scalings of the hysteresis and offers new insights into finite-time/finite-size effect interplay in non-equilibrium systems.
\end{abstract}

\maketitle
\textit{Introduction}.---
Hysteresis is ubiquitous across natural and engineered systems, spanning micro- to macro-scales. This phenomenon manifests in diverse fields, including magnetic and mechanical materials~\cite{bertottiHysteresisMagnetismPhysicists1998,morrisWhatHysteresis2012,wangModeldrivenDatadrivenReview2023}, electric and optical devices~\cite{bertottiScienceHysteresis3volume2005,wangHysteresisElectronicTransport2010,rodriguezProbingDissipativePhase2017}, thermal systems~\cite{fan1995,kuang2000,yildiz2004,wang_scaling_2011}, biological cycles~\cite{nooriHysteresisPhenomenaBiology2014}, chemical reactions~\cite{flanaganHysteresisSolidState1995}, and even economies~\cite{gockeVariousConceptsHysteresis2002}. Characterized by history-dependent dynamics, hysteresis reveals a system's delayed response to external perturbations, where its current state relies on both instantaneous driving parameters and their temporal evolution. Within the realm of physics, in particular, magnetic hysteresis has garnered country-long attention due to its rich phenomenology, with extensive experimental and theoretical investigations~\cite{ewingExperimentalResearchesMagnetism1885,Brokate1996,Mayergoyz2003,tangAnomalousHallHysteresis2016,goodwin2017molecular,moreeReviewHysteresisModels2023}. It reveals fundamental properties like anisotropy, metastable states, coercivity, and remanence, which underpin advancements in magnetic technologies, including permanent-magnet devices~\cite{furlani2001permanent} and magnetic random access memories~\cite{brataas2012current}.  

\begin{figure*}
    \includegraphics{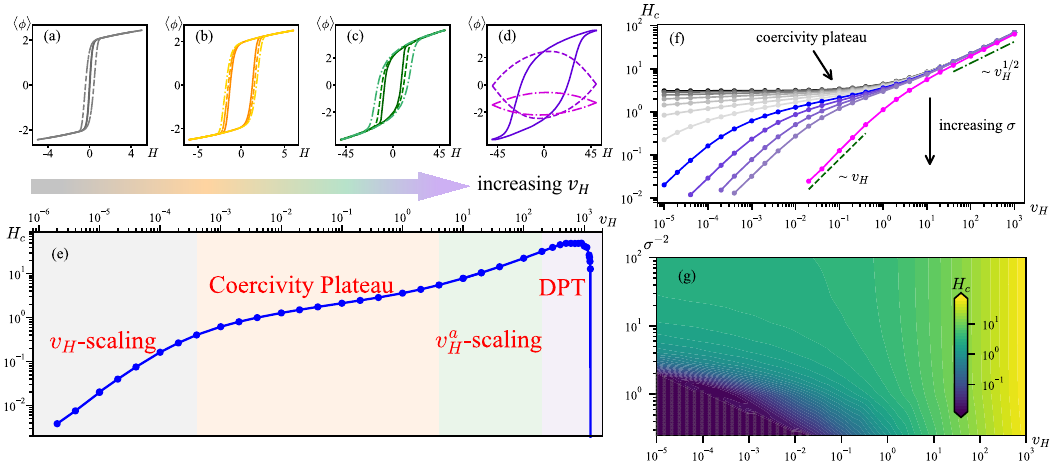}
    \caption{Hysteresis loops and coercivity landscape of the stochastic $\phi^4$ model. Hysteresis loops at (a): $v_H=10^{-5}$(solid), $4\times10^{-4}$(dashed); (b): $v_H=0.01$(solid), $0.02$(dashed), $0.04$(dash-dotted); (c): $v_H=10$(solid), $20$(dashed), $40$(dash-dotted); (d): $v_H=100$(solid), $500$(dashed), $1500$(dash-dotted). (e) Coercivity $H_c$ as a function of driving rate $v_H$ with $\sigma=0.7$, DPT represents dynamic phase transition. (f)~Coercivity landscape $H_c(v_H,\sigma)$, where the noise strength from top to bottom is $\sigma=$ 0, 0.1, 0.2, 0.3, 0.4, 0.5, 0.6, 0.7, 0.8, 0.9, 1.0, 2.0. (g) A contour illustration of (f) after interpolation. In this plot, limited driving amplitude $H_m=50$ is applied in (a-e) while (f-g) not limited driving amplitude to ensure that the order parameter is always saturated~\cite{saturated}; $a_2 = -4$, and all quantities are nondimensionalized by $a_4$ and $\lambda$. }
    \label{fig:phi4-combined}
\end{figure*}

Notably, hysteresis loops exhibit diverse morphologies—some display steep characteristics indicative of first-order phase transitions (FOPT), while others feature smoother, thinner profiles~\cite{lapshinAnalyticalModelApproximation1995,coeyHardMagneticMaterials2011,silveyraSoftMagneticMaterials2018}. A central focus in hysteresis research lies in how hysteresis loops evolve with driving parameters, for example, the rate-dependent deformations in Fig.~\ref{fig:phi4-combined}(a-d). This issue encompasses not only the scaling laws of the hysteresis loop area with dynamic parameters~\cite{jungScalingLawDynamical1990,goldszteinDynamicalHysteresisStatic1997,chakrabartiDynamicTransitionsHysteresis1999a,punyaFrequencyDependenceIsing2010,Wongdamnern2009,Pan2003,Pan2003b} but also delves into phenomena like the dynamic phase transitions (DPT)~\cite{tomeDynamicPhaseTransition1990,chakrabartiDynamicTransitionsHysteresis1999a,jangHysteresisDynamicPhase2001,riegoUnderstandingDynamicPhase2018}. Critical scaling analyses reveal a fundamental dichotomy: systems with and without quasi-static hysteresis exhibit distinct dynamical behaviors~\cite{jungScalingLawDynamical1990,goldszteinDynamicalHysteresisStatic1997}, rooted in contrasting ergodic properties~\cite{venkataramanCrystallineStateEmerging1989,wuErgodicityBreakingScaling2025}. This disparity underscores the interplay between the system's relaxation capability and driving dynamics~\cite{mahatoHysteresisRateCompetition1992,rikvoldMetastableLifetimesKinetic1994,moriAsymptoticFormsScaling2010}. While mean-field theories predict a $2/3$ power law scaling for finite-time deviation from the quasi-static hysteresis in slow-driving FOPT systems~\cite{jungScalingLawDynamical1990,luseDiscontinuousScalingHysteresis1994}, experiments and fluctuating phase-transition models show a richer diversity of scaling behaviors~\cite{he1993observation,acharyyaResponseIsingSystems1995,zhongfanScalingHysteresisIsing1995,acharyyaComparisonMeanfieldMonte1998,liuDynamicHysteresisFerroelectric2002,bar2018kinetic,shuklaHysteresisIsingModel2018,kunduDynamicHysteresisNoisy2023,banerjeeFinitedimensionalSignatureSpinodal2023,rohrleDynamicHysteresisDissipative2024}.

Despite these significant advancements, current studies on hysteresis confront three critical challenges: i) In microscopic theories rooted in nonequilibrium statistical physics, scaling relations of dynamic hysteresis across diverse time scales remain largely unexplored~\cite{jungScalingLawDynamical1990,raoMagneticHysteresisTwo1990,dharHysteresisSelforganizedCriticality1992,luseDiscontinuousScalingHysteresis1994,zhongCompleteUniversalScaling2024,wuErgodicityBreakingScaling2025}; ii) For phenomenological models in magnetism, such as the Jiles-Atherton model~\cite{jilesTheoryFerromagneticHysteresis1986,jiles1992numerical}, Preisach model
~\cite{preisachUeberMagnetischeNachwirkung1935,moreeReviewPlayPreisach2023} 
and Stoner-Wohlfarth model~\cite{Stoner1948AMO}, there is a significant lack of understanding of the continuous transition between quasi-static and dynamic hysteresis~\cite{jilesTheoryFerromagneticHysteresis1986,harrisonPhysicalModelSpin2003}. This raises a critical question: do hysteresis loops have a well-defined quasi-static limit, and how does it emerge with decreasing driving rate? iii) The comparison between experimentally measured and theoretically predicted scaling laws for hysteresis suffers from ambiguity~\cite{yildiz2004,Wongdamnern2009,lee2016scaling,kunduDynamicHysteresisNoisy2023}, primarily due to incomplete characterization of hysteresis dependence on driving rate across the entire accessible frequency regimes.

To address these gaps, we introduce the concept of \textit{coercivity landscape} [Fig.~\ref{fig:phi4-combined}(e)], derived from the driven stochastic $\phi^4$ model~\cite{tuckerOnsetSuperconductivityOneDimensional1971,raoMagneticHysteresisTwo1990,kunduDynamicHysteresisNoisy2023}. This framework offers a unified lens to reconcile scaling relations across time scales and system sizes, highlighting competition between thermodynamic and quasi-static limits in interacting systems. Concomitantly, the derived order parameter's evolution equation serves as a natural bridge between microscopic dynamics and macroscopic hysteretic behaviors, providing a framework for reconciling long-standing theory-experiment mismatches in hysteresis modeling. 

\textit{Coercivity plateau in periodically driven stochastic $\phi^4$ model}.---For a system of interest described by the mean-field Landau model interacting with an external field $H$, its free energy density, as a function of the dimensionless order parameter $\phi$, reads~\cite{reichlModernCourseStatistical2016}
\begin{equation}
\label{eq:phi4-free-energy}
    f_4(\phi, H) = \frac{1}{2} a_2 \phi ^2 + \frac{1}{4} a_4 \phi^4 - H \phi,
\end{equation}
where the constants $a_2$ and $a_4$ encode the reduced temperature and intrinsic interaction strength, respectively. For $a_2<0$ at zero field ($H=0$), the free energy landscape exhibits $\mathbb{Z}_2$-symmetric bistability. Under weak applied fields satisfying $|H| < |H_{\mathrm{sp}}^{\pm}|$, this double-well potential becomes tilted while maintaining both local minima. Here, $H_{\mathrm{sp}}^{\pm} \equiv \pm H^* \equiv \pm \sqrt{(4/27) (-a_2^3/a_4)}$ denotes the spinodal field, at which the metastable state ceases to exist and the metastable minimum merges with the unstable maximum at $\phi_{\mathrm{sp}}^{\pm} \equiv \pm \phi^* \equiv \mp\sqrt{- a_2/(3a_4)}$.
The dynamics of $\phi = \phi(t)$ under a time-dependent field $H = H(t)$ is governed by the Langevin equation~\cite{tuckerOnsetSuperconductivityOneDimensional1971,raoMagneticHysteresisTwo1990,gongFinitetimeScalingLinear2010,kunduDynamicHysteresisNoisy2023}
\begin{equation}
\label{eq:phi4-Langevin}
    \frac{\partial \phi}{\partial t} = -\lambda \frac{\partial f_4(\phi, H)}{\partial \phi} +\zeta(t), 
\end{equation}
where $\lambda$ sets a reference timescale and $\zeta(t)$ is a Gaussian white noise with amplitude $\sqrt{2\lambda\sigma^2}$ satisfying $\langle \zeta (t) \rangle=0$ and $\langle\zeta(t)\zeta(t^{\prime})\rangle=2\lambda\sigma^2\delta(t-t^{\prime})$, and $\sigma \ge 0$ denoting the noise strength. The system defined by Eqs.~\eqref{eq:phi4-free-energy} and \eqref{eq:phi4-Langevin} is referred to as stochastic $\phi^4$ model throughout this work. In the deterministic limit ($\sigma \to 0$), the model describes a sharp FOPT~\cite{jungScalingLawDynamical1990,zhongCompleteUniversalScaling2024}, exactly occurring at the spinodal fields $H = H_{\mathrm{sp}}^{\pm}$ under quasi-static driving from an ordered state to another. Finite driving rates delay the transition beyond $H = H_{\mathrm{sp}}^{\pm}$ due to dynamic hysteresis~\cite{jungScalingLawDynamical1990}. Finite fluctuations ($\sigma > 0$), representing deviations from the thermodynamic limit in our framework~\cite{sigma_N_T,berglundNoiseInducedPhenomenaSlowFast2006,chenMeasurementNoiseMaximumSignature2007,leffFluctuationsParticleNumber2015,fei2024temperature}, cause premature escape from metastable states before reaching $H = H_{\mathrm{sp}}^{\pm}$, which can be estimated by the Kramers' escape rate~\cite{riskenFokkerPlanckEquationMethods1996,moriAsymptoticFormsScaling2010}. The interplay between finite-time and finite-size effects leads to complex nonequilibrium dynamics.

Figures~\ref{fig:phi4-combined}(a-d) display stable hysteresis loops of the stochastic $\phi^4$ model in four different driving regimes, generated under a field swept triangularly with an amplitude $H_m$ and at a rate $|dH/dt| = \lambda v_H$ (See companion paper~\cite{CPregular} for details). Two key features are observed: i) The loops in (a-c) are approximately centrally symmetric with respect to the origin, and the loop width increases monotonically with $v_H$ when the saturation field~\cite{saturated} is smaller than $H_m$; however, at very fast driving rates, the loops in (d) shrink and form narrow cycles centered near the initial value of the order parameter. ii) In the first three regimes, the hysteresis originates from field-driven switching between the two phases of the underlying FOPT. Such switching remains sharp in the former two regimes under slow driving but gradually becomes rounded in the third regime with fast driving.

In the field-driven transition from one stable phase to another, coercivity ($H_c$) is defined as the applied field value at which the ensemble-averaged order parameter crosses zero~\cite{Hcdefinition}. Figure~\ref{fig:phi4-combined}(e) illustrates a coercivity landscape obtained from the periodically driven stochastic $\phi^4$ model, showing its evolution with increasing driving rate at a fixed field amplitude. This panorama strikingly reveals four distinct dynamic regimes: i) near-equilibrium regime where coercivity scales linearly with $v_H$; ii) a coercivity-plateau regime where $H_c$ is nearly independent of $v_H$; iii) post-plateau power-law scaling regime; and iv) DPT regime characterized by rapidly diminishing coercivity at high driving rates. Unlike the often-studied loop area response curves, which typically exhibit a long, decreasing tail after a maximum with increasing driving frequency~\cite{chakrabartiDynamicTransitionsHysteresis1999a,punyaFrequencyDependenceIsing2010}, this coercivity landscape uniquely focuses on the ascending regime, clearly identifying and explaining its distinct dynamic behaviors. Notably, the existence of the coercivity plateau is a novel finding of this study.

The coercivity plateau's dependence on the noise strength is illustrated in the coercivity landscape $H_c(v_H,\sigma)$[Fig.~\ref{fig:phi4-combined}(f)]~\cite{monodriving}, where $\sigma$ increases from top (deterministic limit $\sigma\to0$) to bottom. As $\sigma$ decreases, the coercivity plateau extends toward the quasi-static limit ($v_H\to0$), consisting with experimental observation~\cite{lee_magnetization_1999}; conversely increasing $\sigma$ narrows the plateau until it vanishes, squeezed by adjacent regimes. This plateau arises from the competition between the lifetime of the metastable state~\cite{binderTimeDependentGinzburgLandauTheory1973} and the variation of the potential landscape. Intriguingly, a non-commutativity of limits emerges at the far left side of the coercivity plateau:
\begin{equation}
    \begin{cases}
    \lim_{\sigma\to0}\lim_{v_H\to 0}H_c = 0\\
    \lim_{v_H \to 0}\lim_{\sigma\to0}H_c = H^*,\label{noncom}
    \end{cases}
\end{equation}
which is clearly illustrated by the contour plot in Fig.~\ref{fig:phi4-combined}(g). Taking the quasi-static limit first implies that finite fluctuations are always present to ensure ergodicity. In contrast, taking the thermodynamic limit first eliminates fluctuations, causing the system to remain trapped in metastable states even under quasi-static driving, rendering the global equilibrium state unreachable. This distinction has also been reflected in the difference between deterministic mean-field results and fluctuating Monte Carlo simulations of the Ising model~\cite{acharyyaComparisonMeanfieldMonte1998}. In this sense, for macroscopic and mesoscopic materials, within the achievable observation time, the quasi-static hysteresis can be well-defined with the coercivity associated with the field-driven ergodicity-breaking FOPT point; for microscopic materials with non-negligible fluctuations, it is expected that the hysteresis loop will disappear during slow field sweeping and approach the equilibrium magnetization curve.

\textit{Finite-time scaling relations}.---To quantify the evolution of the ensemble-averaged order parameter, we derive the corresponding Fokker-Planck equation of Eq.~\eqref{eq:phi4-Langevin} as
\begin{equation}
    \label{eq:phi4-FP-equation}
    \frac{\partial P(\phi, t)}{\partial t} = \lambda \frac{\partial}{\partial \phi}\left[\frac{\partial f_4(\phi,H)}{\partial \phi} P(\phi,t)\right] + \lambda\sigma^2\frac{\partial^2P(\phi,t)}{\partial\phi^2},
\end{equation}
where $P(\phi, t)$ denotes the probability density of the system having the value $\phi$ at time $t$. Multiplying Eq.~\eqref{eq:phi4-FP-equation} by $\phi$ and integrating the resulting expression over all possible values of $\phi$, an ensemble-averaged equation is derived as,
\begin{equation}
\label{eq:phi4-averaged-value-evolution}
    v_H \frac{d\langle\phi\rangle}{dH} = -a_2 \langle \phi \rangle - a_4 \langle \phi^3\rangle + H,
\end{equation}
where $dH=\lambda v_H dt$ has been used. Hereafter, the increasing branch of the field function is considered in analytical calculations, unless specifically stated otherwise. As the first main result of this Letter, Eq.~\eqref{eq:phi4-averaged-value-evolution} captures the evolution of measurable macroscopic quantities and connects microscopic stochastic models with phenomenological equations widely used in magnetism~\cite{moreeReviewHysteresisModels2023}. 

By evaluating Eq.~\eqref{eq:phi4-averaged-value-evolution} at $\langle\phi\rangle=0$, we find the coercivity 
\begin{equation}
\label{eq:phi4-coercivity-expression}
    H_c = v_H \eval{\frac{d\langle \phi \rangle}{d H}}_{\langle \phi \rangle= 0} + a_4\eval{\langle \phi^3\rangle}_{\langle \phi \rangle= 0},
\end{equation}
from which, three distinct scaling regimes for $H_c$ with respect to $v_H$ are identified: the near-equilibrium regime, the post-plateau regime, and the fast-driving regime~\cite{CPregular}. 
In the near-equilibrium regime, for non-divergent equilibrium value of $\eval{d\langle\phi\rangle / dH}_{\langle \phi \rangle =0}$, the scaling 
\begin{equation}
    H_c \sim v_H
\end{equation} 
is denoted by the dashed line in Fig.~\ref{fig:phi4-combined}(f). This linear scaling holds for systems weakly driven out of equilibrium~\cite{esposito2012stochastic,martinez2016brownian,ma2020experimental,li_geodesic_2022,zhaoEngineeringRatchetbasedParticle2024}.
In the fast-driving regime, the dynamics is effectively governed by a passive relaxation process toward the mono-stable minimum induced by the continuously increasing field. The interplay between the relaxation timescale and the field variation rate leads to
\begin{equation}
    H_c \sim v_H^{1/2},
\end{equation}
which is denoted by the dash-dotted line in Fig.~\ref{fig:phi4-combined}(f). While previous analytical studies often overlooked this fast-driving regime, this scaling finds support in existing numerical~\cite{zhongDynamicScalingHysteresis1994,zhuDynamicsScalingLowfrequency2007} and experimental evidence~\cite{nistorMagneticEnergyLoss2005}.

\begin{figure}
    \centering
    \includegraphics{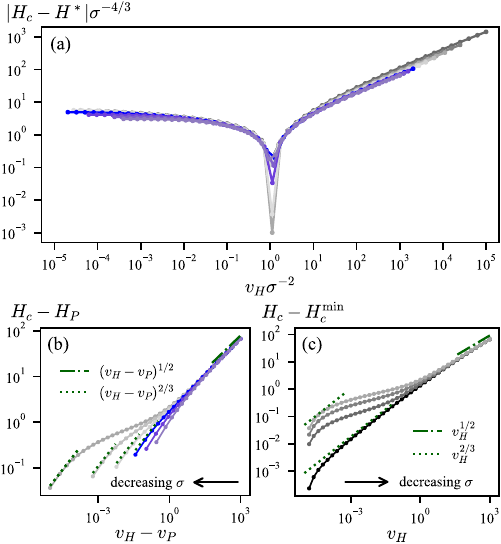}
    \caption{(a) Finite-size scaling of the coercivity plateau. $\sigma$ from 0.1 to 1.0 with interval 0.1. (b) Coercivity reduced by the reference plateau coercivity. Curves from right to left correspond to noise strengths $\sigma =$ 1.0, 0.8, 0.7, 0.5, 0.4, 0.3. (c) Coercivity reduced by the minimum observed coercivity within the simulation timescale shown in Fig.~\ref{fig:phi4-combined}(f). Curves from left to right correspond to $\sigma =$ 0.3, 0.2, 0.1, 0.} 
    \label{fig:phi4-coercivity-plateau-and-after}
\end{figure}

\textit{Finite-size scaling of the coercivity plateau}.---Before analyzing the finite-time behavior of coercivity near the plateau, we first examine the plateau itself. Expanding the free energy~\eqref{eq:phi4-free-energy} near the spinodal point $(H^*,\phi^*)$ and retaining terms up to third order, the resulting discretized equation governing the variation of the order parameter is
\begin{equation}
\label{eq:phi4-Langevin-increment-phi3}
    \delta \varphi = - a_3 \varphi^2 v_H^{-1} \delta h + h v_H^{-1} \delta h + \sqrt{2 \sigma^2 v_H^{-1} \delta h}\;W,
\end{equation}
where $\varphi\equiv\phi-\phi^{*}$, $h\equiv H-H^{*}$ and $a_3 \equiv - \sqrt{-3 a_2 a_4}$. To uncover the finite-size (noise) effects encoded in the coercivity plateau, we employ a scaling collapse method inspired by renormalization group theory~\cite{goldenfeldLecturesPhaseTransitions2018,zhongCompleteUniversalScaling2024}. Using $\Sigma$ as the scaling factor to rescale the noise strength from $\sigma$ to $\sigma^{\prime} = \sigma \Sigma^{-1}$, any quantity $\mathcal{O}$ transforms as $\mathcal{O}^{\prime} = \mathcal{O} \Sigma^{-[\mathcal{O}]}$, where $[\mathcal{O}]$ is the scaling dimension of $\mathcal{O}$. To preserve the form of the dynamical equation under scaling transformations, the scaling dimensions of all terms in Eq.~\eqref{eq:phi4-Langevin-increment-phi3} must be consistent. By setting the scaling dimension of the coefficient of the highest-order term to be zero, i.e., $[a_3]=0$~\cite{zhongCompleteUniversalScaling2024}, we obtain $[h]=4/3$ and $[v_H]=2$. Therefore, the second main result of this Letter is obtained as~\cite{CPregular} 
\begin{equation}
\label{eq:phi4-finite-size-scaling-plateau}
    H^* - H_{P} \sim \sigma^{4/3} \quad \text{at} \quad v_{P} \sim \sigma^2,
\end{equation}
where $H_{P}$ denotes the height of the coercivity plateau ($\lim_{\sigma\to0}H_P = H^*$), and $v_{P}$ is the corresponding reference driving rate~\cite{Vp}. By rescaling the axes in Fig.~\ref{fig:phi4-combined}(f) according to Eq.~\eqref{eq:phi4-finite-size-scaling-plateau}, it is found that the coercivity curves for small $\sigma\in[0.1,1]$ approximately collapse onto a single universal curve [Fig.~\ref{fig:phi4-coercivity-plateau-and-after}(a)], confirming the validity of the finite-size scaling relations. The departure from the scaling behavior at both ends can be traced to the disappearance of the coercivity plateau in the near-equilibrium and fast-driving regimes. This curve diverges at a critical driving rate that yields $H_c = H^*$. In the $\sigma \to 0$ limit, this critical driving rate approaches zero, rendering the region to its left inaccessible. This explains why the coercivity plateau—arising from taking the thermodynamic limit first in Eq.~\eqref{noncom}—does not vanish in the slow-driving regime. In contrast, for finite-size systems, coercivity values span both sides of this critical point, with the gently varying region at lower driving rates reflecting the presence of a coercivity plateau.

\textit{The $2/3$ power law scaling.}---Adopting a procedure analogous to the finite-time scaling analysis in the thermodynamic limit~\cite{jungScalingLawDynamical1990,zhongCompleteUniversalScaling2024}, we expand Eq.~\eqref{eq:phi4-FP-equation} around the coercivity plateau point $(v_P,H_P)$ and reveal~\cite{CPregular}
\begin{equation}
    H_c - H_P \sim (v_H - v_P)^{2/3}.
\end{equation}
In the deterministic limit ($\sigma \to 0$), $(H_P,v_P)\rightarrow(H^*,0)$ as a result of Eq.~\eqref{eq:phi4-finite-size-scaling-plateau}, and thus $H_c - H^* \sim v_H^{2/3}$. This is consistent with the relation $A-A_0\sim v_H^{2/3}$, where $A$ ($A_0$) is the dynamic (quasistatic) hysteresis loop area~\cite{jungScalingLawDynamical1990,luseDiscontinuousScalingHysteresis1994,wuErgodicityBreakingScaling2025}. Such consistency is due to the fact that the hysteresis loop in this regime [Fig.~\ref{fig:phi4-combined}(b)] is geometrically approximated to a parallelogram with $A-A_0 \propto H_c - H^*$. The $2/3$-exponent scaling for finite $\sigma$ is demonstrated by the dotted lines in Fig.~\ref{fig:phi4-coercivity-plateau-and-after}(b). For very small $\sigma$, the genuine plateau point becomes unreachable within accessible simulation timescales [upper set of Fig.~\ref{fig:phi4-combined}(f)], making it challenging to characterize the coercivity plateau. Consequently, it is practical to take the minimum observed coercivity $(H_{c}^{\mathrm{min}})$ as a quasi-static value. Figure~\ref{fig:phi4-coercivity-plateau-and-after}(c) illustrates the scaling of $H_c-H_c^{\mathrm{min}}$ with $v_H$, revealing a transient $2/3$-exponent scaling that agrees with observations in Ref.~\cite{wuErgodicityBreakingScaling2025}. In this case, the accuracy of the scaling exponent obtained experimentally will decrease as the system becomes smaller, which likely causes deviations from the mean-field results. In this sense, the coercivity landscape across multi-driving regimes offers a panoramic view of rich hysteresis scalings and serves as an effective experimental reference.

\begin{figure}
    \centering
    \includegraphics{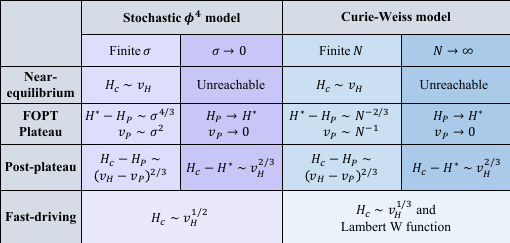}
    \caption{Finite-time and finite-size scalings for hysteresis in interacting systems. The results for the Curie-Weiss model of $N$ spins are obtained in the companion paper~\cite{CPregular}.}
    \label{fig:summary-finite-time-scaling}
\end{figure}

\textit{Concluding remarks}.---The coercivity landscape is proposed to characterize the hysteresis phenomena across timescales. The coercivity plateau reveals the interplay between finite-time and finite-size effects in FOPT, unifying diverse findings within equilibrium~\cite{fisherScalingFirstorderPhase1982,binderFinitesizeScalingFirstorder1984} and nonequilibrium~\cite{gongFinitetimeScalingLinear2010} frameworks into a coherent picture. Analyzing hysteresis beyond physical models, such as in climate and neural dynamics~\cite{berglundNoiseInducedPhenomenaSlowFast2006}, with our approach holds promise. We summarize our main findings in Fig.~\ref{fig:summary-finite-time-scaling}, where the Curie-Weiss model is investigated in the companion paper~\cite{CPregular}. The coercivity plateau bridges the gap between irreversibility in the linear response regime and irreversibility with signatures of ergodicity breaking, reflecting the competition between the thermodynamic and quasi-static limits. Remarkably, the scaling relations near the coercivity plateau are universal across both the stochastic $\phi^4$ and Curie-Weiss models~\cite{systems}. However, these two models exhibit different scaling behaviors in the fast-driving regime, where the system transitions from the finite-time FOPT to the DPT regime. The discrepancy in this regime highlights the sensitivity of far-from-equilibrium dynamics to model-specific details, and thus different microscopic~\cite{raoMagneticHysteresisTwo1990,shuklaHysteresisIsingModel2018} and mathematical models~\cite{berglundNoiseInducedPhenomenaSlowFast2006} may contain richer scalings. 

We end this Letter with three remarks: i) Currently, the scaling exponents of hysteresis loops measured experimentally in different systems vary, and there are ambiguities in their theoretical explanations (See End Matter for details). The coercivity landscape provides a reliable basis for measuring the dynamic scaling of hysteresis. The finite-size scaling of coercivity is expected to be verified using microscopic experimental platforms such as cold-atom systems~\cite{lee2016scaling,banerjeeFinitedimensionalSignatureSpinodal2023} or two-dimensional ferromagnetic materials~\cite{mengAnomalousThicknessDependence2021,yiTwodimensionalAnionrichNaCl22025}. ii) The phenomenological hysteresis models in the realm of magnetism and the research on the dynamics of FOPT in statistical physics previously seemed to be two independent areas. This work establishes a connection between them and presents a theoretical framework for coercivity based on mean-field theory. Besides the coercivity itself, the \textit{susceptibility} at the coercive point, $\chi_c\equiv \eval{d\langle \phi \rangle/d H}_{H=H_c}$, is another crucial characteristic quantity~\cite{binderFinitesizeScalingFirstorder1984,jiles1992numerical,milleProbingDynamicsNanoparticle2021}. According to Eq.~\eqref{eq:phi4-coercivity-expression}, the finite-time and finite-size scalings of $\chi_c$ can be naturally obtained from the coercivity landscape. iii) The dynamics of FOPT and DPT can be uniformly characterized by coercivity~\cite{Dynamicphase}. On this basis, the finite-time thermodynamics of interacting systems with phase transitions driven at different timescales can be systematically studied. Previous studies~\cite{sivakThermodynamicMetricsOptimal2012,maOptimalOperatingProtocol2018,ma2020experimental,li_geodesic_2022,zhaoEngineeringRatchetbasedParticle2024} in the near-equilibrium regime have revealed that different driving protocols, although not affecting the $v_H$-scaling, can influence irreversibility by changing the scaling coefficient. In this work, we use a linear field sweep to determine the finite-time scalings of coercivity. Exploring the optimal control strategies for coercivity and the corresponding energy cost in regimes beyond $v_H$-scaling is of great significance for the optimization of phase transition systems~\cite{meibohmFiniteTimeDynamicalPhase2022,wuErgodicityBreakingScaling2025}. Relatedly, the performance of thermal machines with phase transition~\cite{campisiPowerCriticalHeat2016,maQuantumThermodynamicCycle2017,liangMinimalModelCarnot2025a}, and the information thermodynamics of interacting systems~\cite{rolandiCollectiveAdvantagesFiniteTime2023,lipka2024minimizing} are ‌fascinating issues worthy of further exploration. Exploring the relation between finite-time scaling in field-driven systems and nonequilibrium relaxation dynamics is also a promising direction for future research, particularly in view of the successful application of renormalization-group theory in both contexts~\cite{loscarSpinodalsCriticalPoint2016,meibohmFiniteTimeDynamicalPhase2022,zhongCompleteUniversalScaling2024}.

\textit{Acknowledgments}.---We thank Sai Li for the helpful discussions in the early stage of this work. We would like to express our sincere gratitude to the anonymous referees for their constructive comments. This work is supported by the National Natural Science Foundation of China under grant No. 12305037 and the Fundamental Research Funds for the Central Universities under grant No. 2023NTST017.

\textit{Data availability}.---No data were created or analyzed in this study.

\bibliography{mainref}

\begin{thebibliography}{114}
\bibitem [{saturated()}]{saturated}%
  \BibitemOpen
  \bibinfo {note} {“saturated” or “saturation” used in this study refers to hysteresis loops for which the irreversibility (loop area) no longer increases upon further increasing the field amplitude. Operationally, under periodic driving starting from an ordered state, a saturated hysteresis loop becomes stable after one driving period, whereas an unsaturated loop remains in a transient state during the early driving periods.}\BibitemShut {Stop}%
\bibitem [{Hcdefinition()}]{Hcdefinition}%
  \BibitemOpen
  \bibinfo {note} {For symmetric hysteresis loops, the coercivity intersections corresponding to the increasing and decreasing branches of the field are equal in magnitude. However, in the fast-driving dynamic-phase-transition regime, these two intersection points can differ. To account for both cases, a universal definition of coercivity is adopted in this study: $H_c\equiv(|H_{c1}|+|H_{c2}|)/2$, where $H_{c1}$($H_{c2}$) is the first (second) intersection point of the hysteresis loop with the horizontal line $\langle \phi \rangle =0$.}\BibitemShut {Stop}%
\bibitem [{monodriving()}]{monodriving}%
  \BibitemOpen
  \bibinfo {note} {In Fig.~\ref{fig:phi4-combined}(f), for simulation simplicity, we adopt monotonically increasing-field driving instead of periodic driving. The initial field magnitude is chosen slightly larger than the spinodal field for the given $a_{2}$ and $a_{4}$, such that the free-energy landscape has a single stable minimum, and the system is prepared in the corresponding equilibrium state.}\BibitemShut {Stop}%
\bibitem [{vp()}]{Vp}%
  \BibitemOpen
  \bibinfo {note} {A specific operational definition for reference plateau point ($v_P$) is the driving velocity at which the $H_c(v_H)$ curve exhibits a minimal logarithmic derivative with respect to $v_H$ in the log-log plots~\cite{CPregular}}\BibitemShut {NoStop}%
\end{thebibliography}

\onecolumngrid{}

\begin{center}
\textbf{End Matter}
\end{center}

\twocolumngrid{}
\begin{table*}[!htb]
\caption{\label{tab:Experiments} Experimental Results of Hysteresis Scaling}

\begin{tcolorbox}[colback=blue!3!white,colframe=black,arc=5mm,opacityback=1]
\centering
\renewcommand{\arraystretch}{1.5}
\scriptsize
\setlength{\tabcolsep}{6pt}

\begin{ruledtabular}

\begin{tabular}{p{0.45\textwidth}|p{0.30\textwidth}|p{0.20\textwidth}}
\noalign{\vskip 3pt}
\multicolumn{3}{c}{
{\footnotesize   
\parbox{\linewidth}{\centering
\textcolor{blue}{\textbf{Total Loop Area Scaling}: $A\propto v_H^{\alpha}H_m^{\beta}$}
\parbox[c]{0.52\linewidth}{\centering
Statistical Theory~\cite{raoMagneticHysteresisTwo1990,acharyyaResponseIsingSystems1995}\\
Phenomenological Model~\cite{jieli_li_improved_2001,venkatachalam_accurate_2002}}
}}
}\\
\noalign{\vskip 3.5pt}
\hline\hline

\cA{\textbf{System}} &
\cB{\textbf{Scaling Exponents}} &
\cC{\textbf{Hysteresis Type}} \\
\hline

\cA{PbTiO\textsubscript{3}/polymer ferroelectric composites~\cite{Pan2003b}} &
\cB{$\alpha=1/3,~\beta=2/3$} &
\cC{\multirow{7}{*}{Field-Driven}} \\
\cline{1-2}

\cA{Ferroelectric SrBi\textsubscript{2}Ta\textsubscript{2}O\textsubscript{9} thin films~\cite{Pan2003}} &
\cB{$\alpha=2/3,~\beta=2/3$} &
\cC{} \\
\cline{1-2}

\cA{Ferroelectric BaTiO\textsubscript{3} single crystals~\cite{Wongdamnern2009}} &
\cB{$\alpha=1.667H_m-2.804,~\beta=4.157$} &
\cC{} \\
\cline{1-2}

\cA{Ferroelectric BaTiO\textsubscript{3} bulk ceramics~\cite{wongdamnern2010hysteresis}} &
\cB{$\alpha=-0.39,~\beta=1.06$} &
\cC{} \\
\cline{1-2}

\cA{Soft Pb(Ti, Zr)O\textsubscript{3} ferroelectric ceramic~\cite{yimnirun2007dynamic_soft}} &
\cB{$\alpha=-0.25,~\beta=0.89$} &
\cC{} \\
\cline{1-2}

\cA{Hard Pb(Ti, Zr)O\textsubscript{3} ferroelectric ceramic~\cite{yimnirun2007dynamic_hard}} &
\cB{$\alpha=-0.28,~\beta=0.89$} &
\cC{} \\
\cline{1-2}

\cA{Ultrathin Fe/Au ferromagnetic film~\cite{he1993observation}} &
\cB{$\alpha=0.31\pm0.05,~\beta=0.59\pm0.07$} &
\cC{} \\
\hline

\cA{Structural transition in VO\textsubscript{2}~\cite{zhangScanningrateDependenceEnergy1996}} &
\cB{$\alpha=1/2,~\beta=0$} &
\cC{Temperature-Driven} \\

\hline
\noalign{\vskip 13pt}
\hline
\hline

\noalign{\vskip 4pt}
\multicolumn{3}{c}{
{\footnotesize
\parbox{\linewidth}{\centering
\textcolor{blue}{\textbf{Excess Loop Area Scaling}: $A-A_0\propto v_H^{n}$}
\parbox[c]{0.52\linewidth}{\centering
Statistical Theory~\cite{jungScalingLawDynamical1990,luseDiscontinuousScalingHysteresis1994}\\
Phenomenological Model~\cite{colaiori_loss_2006,bertotti_hysteresis_1998}}
}}}\\\noalign{\vskip 4pt}
\hline\hline

\cA{\textbf{System}} &
\cB{\textbf{Scaling Exponent}} &
\cC{\textbf{Hysteresis Type}} \\
\hline

\cA{Evaporated permalloy thin film~\cite{nistorMagneticEnergyLoss2005}} &
\cB{$n=0.49\pm0.01$} &
\cC{\multirow{3}{*}{Field-Driven}} \\
\cline{1-2}

\cA{Cold atomic system~\cite{lee2016scaling}} &
\cB{$n=0.64\pm0.04$} &
\cC{} \\
\cline{1-2}

\cA{Switching of bistable laser~\cite{jungScalingLawDynamical1990}} &
\cB{$n=2/3$} &
\cC{} \\
\hline

\cA{Mott transition in V\textsubscript{2}O\textsubscript{3}~\cite{bar2018kinetic}} &
\cB{\begin{tabular}[c]{@{}c@{}}$n=0.62\pm0.06$ (heating)\\$n=0.64\pm0.09$ (cooling)\end{tabular}} &
\cC{\multirow{7}{*}{Temperature-Driven}} \\
\cline{1-2}

\cA{MnNiSn based Heusler alloy~\cite{bar_suppression_2021}} &
\cB{\begin{tabular}[c]{@{}c@{}}$n=0.93\pm0.13$ (heating)\\$n=0.85\pm0.07$ (cooling)\end{tabular}} &
\cC{} \\
\cline{1-2}

\cA{N-SmA transition in binary mixture~\cite{yildiz2004}} &
\cB{\begin{tabular}[c]{@{}c@{}}$n=0.672\pm0.008$ (mixture ratio $4:6$)\\$n=0.766\pm0.05$ (mixture ratio $3:7$)\\$n=0.701\pm0.04$ (mixture ratio $2:8$)\\$n=0.629\pm0.005$ (mixture ratio $1:9$)\end{tabular}} &
\cC{} \\

\end{tabular}

\end{ruledtabular}
\end{tcolorbox}
\end{table*}

\textit{Experimental results on the rate-dependence of hysteresis}---. Table~\ref{tab:Experiments} lists rich dynamic scalings of hysteresis observed in experiments, where the scaling of the total hysteresis loop area ($A\propto v_H^{\alpha}H_m^{\beta}$) and the scaling of the excess hysteresis loop area ($A-A_0\propto v_H^{n}$) generally depend on different theoretical explanations. These two types of scaling may appear in different parameter regions and can be related through the coercivity landscape, as we have shown in Fig.~\ref{fig:summary-finite-time-scaling}. Particularly, $A-A_0\propto v_H^{2/3}$, as a universal prediction of mean-field theories in both field-driven~\cite{jungScalingLawDynamical1990,luseDiscontinuousScalingHysteresis1994} and temperature-driven~\cite{zhengThermalHysteresisScaling1998} systems, finds support in systems such as the switching of bistable laser~\cite{jungScalingLawDynamical1990} and the Mott transition in $V_2O_3$~\cite{bar2018kinetic}. However, to the best of our knowledge, the experimentally observed exponents near 1/3~\cite{Pan2003b,he1993observation} and 1/2~\cite{nistorMagneticEnergyLoss2005} have not yet been robustly explained by existing theories~\cite{raoMagneticHysteresisTwo1990,dharHysteresisSelforganizedCriticality1992}, and our current results provide a potential account for them. Besides these typical scalings, a broader range of exponent behavior has been observed across systems, especially since many scalings break down outside a narrow frequency range~\cite{wongdamnern2010hysteresis,yimnirun2007dynamic_soft,yimnirun2007dynamic_hard,lee2016scaling,yildiz2004,kuang2000,bar_suppression_2021,wang_scaling_2011}. For example, in certain ferroelectric ceramics~\cite{wongdamnern2010hysteresis,yimnirun2007dynamic_soft,yimnirun2007dynamic_hard}, the measurements fall in the high-frequency regime and exhibit negative exponents. The difference in the exponents stems from variations in amplitude $H_m$ and the degree of saturation at different $v_H$ or $H_m$. 

Moreover, the ambiguity in defining the “quasi-static” plateau height ($H_P$ or $A_0$) systematically distorts the fitted exponent. In practice, the true plateau is often unreachable; instead, it is commonly approximated by either the minimal observed coercivity or an arbitrarily chosen fitting constant. As a result, the thus-determined $A_0$, which deviates from the true quasi-static plateau, gives rise to an extremely narrow frequency range where the 2/3 scaling emerges, as illustrated in Fig.~\ref{fig:phi4-coercivity-plateau-and-after}(c). Compounding this issue, the driving frequency in experiments may not strictly meet the requirements of the slow-driving FOPT regime. Together, these factors collectively lead to either the absence of stable 2/3 scaling behavior in experiments or the acquisition of scaling exponents with significant deviations from 2/3 during data fitting.

Different from the concerns of statistical physics regarding the dynamic scaling of hysteresis, phenomenological models in magnetism, such as the Jiles-Atherton model~\cite{jilesTheoryFerromagneticHysteresis1986,jiles1992numerical}, Preisach model~\cite{preisachUeberMagnetischeNachwirkung1935,moreeReviewPlayPreisach2023}, and Stoner-Wohlfarth model~\cite{Stoner1948AMO}, focus primarily on fitting actual measured hysteresis curves to determine the material-specific empirical parameters. Their dynamic extensions can indeed yield apparent driving-rate scalings once additional kinetic ingredients are introduced, typically formulated as the rate-dependent excess contribution (e.g. $H_c-H_P$). Some phenomenological models lead to scaling crossovers across different driving-rate regimes~\cite{colaiori_loss_2006}, which agrees with our findings and Ref.~\cite{wuErgodicityBreakingScaling2025}. Besides the power law scaling, log-type scaling behavior have also been reported in alternative scenarios~\cite{sharrock_time_1994}. However, such scalings are typically governed by phenomenologically introduced dynamical parameters and the assumed loss mechanisms built into the model, and the corresponding dynamic extensions are therefore often geared toward engineering prediction rather than multi-regime panoramic scaling. In this regard, our statistical framework may help to systematically identify the key dynamical ingredients and to clarify when and why such scaling behaviors emerge.  Further possible explorations include establishing the basis of the empirical model starting from the order parameter evolution equation we obtained [Eq. \eqref{eq:phi4-averaged-value-evolution}]; or starting from the empirical model, studying the dynamic scaling of hysteresis and comparing it with the statistical model and experiments.

It is also worth noting that the results of this Letter and the companion paper~\cite{CPregular} indicate that, in the slow-driving regime, the hysteresis of small systems is more sensitive to driving dynamics, whereas large systems converge to quasi-static hysteresis. Consequently, when measuring characteristic quantities, such as anomalous Hall conductance~\cite{tangAnomalousHallHysteresis2016,kayyalha_absence_2020}, in mesoscopic or microscopic systems via hysteresis loops, one must carefully distinguish between the statistical characteristics in observables and the dynamically dependent features arising from the choice of driving rate. Otherwise, a gap may emerge between experimental results and relevant theories, making the former unsuitable as direct validation evidence. The coercivity landscape we present can serve as a useful reference for such analyses.

Finally, we leave two remarks here: i) the real systems listed in Table~\ref{tab:Experiments} typically require nucleation and growth to undergo a first-order phase transition. Therefore, the specific dynamic scaling exponents in different driving rate regimes may differ from the results we obtained based on the mean-field model. We stress that the primary aim of this Letter is to provide a concise and intuitive framework for the hysteresis phenomenon, with the goal of inspiring the exploration of richer hysteresis scaling in real experimental systems. Without a panoramic view of hysteresis, researchers have been limited to using theories with different types of scaling forms within localized parameter regions to process their experimental results. By contrast, our coercivity landscape maps all data onto a single, unified control variable, thereby may eliminate system-specific biases and reconcile these inconsistent findings. ii) We note that a recent theoretical study~\cite{zhongCompleteUniversalScaling2024} provides an elegant framework for constructing complete universal scaling applicable across different driving regimes by co-varying system parameters, such that hysteresis curves at different driving conditions collapse to a unique one. The diverse scaling behaviors reported in this Letter, which emerge in different driving regimes under fixed system parameters, do not contradict this complete scaling; we elaborate on this consistency in the companion paper~\cite{CPregular}. Moreover, our work bridges Ref.~\cite{zhongCompleteUniversalScaling2024} and experimental platforms, potentially inspiring verification of the complete scaling theory for finite-time FOPT hysteresis by identifying critical parameter regions that exhibit distinct dynamical behaviors in experiments.

\end{document}